\documentclass[graybox]{svmult}

\usepackage{type1cm}        
\usepackage{cite}                            
\usepackage{makeidx}         
\usepackage{graphicx}        
\usepackage{multicol}        
\usepackage[bottom]{footmisc}

\usepackage{newtxtext}       %
\usepackage[varvw]{newtxmath}       
\usepackage[numbers]{natbib}

\makeindex             

\begin{document}

\title*{Can we detect super-Chandrasekhar white dwarfs via continuous gravitational waves?}

%
%
\author{Mayusree Das and Banibrata Mukhopadhyay}
\institute{Mayusree Das \at Joint Astronomy Programme, Department of Physics, Indian Institute of Science, Bangalore 560012, \email{mayusreedas@iisc.ac.in}
\and Banibrata Mukhopadhyay \at Joint Astronomy Programme, Department of Physics, Indian Institute of Science, Bangalore 560012 \email{bm@iisc.ac.in}}
%

\maketitle
\vspace{-2 cm}
    \textit{To be published in Astrophysics and Space Science Proceedings, titled "The Relativistic Universe: From Classical to Quantum, Proceedings of the International Symposium on Recent Developments in Relativistic Astrophysics", Gangtok, December 11-13, 2023: to felicitate Prof. Banibrata Mukhopadhyay on his 50th Birth Anniversary", Editors: S Ghosh \& A R Rao, Springer Nature.}

\vspace{1 cm}

\abstract{ The existence of massive white dwarfs (WDs) containing more than Chandrasekhar's maximum mass has been suggested via the detection of peculiar type Ia Supernovae. It had been crucial to directly detect those 'super' (more massive)-Chandrasekhar WDs to confirm their existence. The WD's small size and cold internal environment have been a great disadvantage for detecting them via electromagnetic surveys like GAIA and SDSS. Although many WDs are detected in optical observations, none of the super-Chandrasekhar WDs was observed directly. Mukhopadhyay and his group proposed a decade ago that one of the possibilities to explain the high mass of super-Chandrasekhar WD is to exercise the existence of a strong magnetic field in the WD interior. The group also recently proposed that a high magnetic field, in turn, can nontrivially deform the star from a spherical shape. In the presence of rotation and obliquity angle, such WDs can radiate continuous gravitational waves (CGW). This opens up the exciting possibility of detecting the super-Chandrasekhar WDs with the upcoming detector LISA. However, the gravitational wave (GW) amplitude will decay with time due to electromagnetic radiation and quadrupolar radiation. Altogether, those decay mechanisms will set a timescale for detecting WDs via CGW. Even with a timescale bound, we can expect to detect a few super-Chandrasekhar WDs with a couple of years of cumulative detection, leading to the direct detection of super-Chandrasekhar WDs. }



\section{Introduction}

White dwarfs (WDs) are the end state of low to moderately mass main sequence stars. The contraction of the main sequence star comes to a halt after the electrons become degenerate under huge density and support against gravitational force. It is famously known that the WDs have a mass limit of $1.4M_\odot$- Chandrasekhar's mass limit. If the WD accretes beyond Chandrasekhar's mass limit, then it will result in a type Ia supernova. However, there are overluminous Type Ia supernovae, e.g. SN 2003fg, had been observed with inferred progenitor's mass is as high as $2.8M_\odot$ \citep{howell2006}, violating Chandrasekhar's mass limit significantly. Mukhopadhyay and his group \citep{das2014} proposed that the excess mass of the 'super'-Chandrasekhar is held by excess pressure generated due to a high magnetic field. Incidentally, the WDs may retain a high magnetic field in their core from their main sequence progenitors, as suggested by \citep{wick2005}, or may be generated during the end of the main sequence phase and carried on to WD by flux freezing. A dynamo operating during binary evolution can also generate the WD's magnetic field if they are in a double WD system \citep{tout2008}. It is also widely known that a magnetic field up to $10^9$G is observed on WD's surface (see for review \citep{Ferrario2015, Ferrario2020}). However, the magnetic field can be four orders of magnitude higher near the core. The presence of a high magnetic field, in turn, can deform the star from its spherical shape. Mukhopadhyay's group proposed that in the presence of rotation along a misaligment between the magnetic and rotation axes, the deformed WD can radiate gravitational waves (GW) \citep{KM2019}. Like binary merger systems, this kind of isolated star also poses a non-zero time-varying quadrupole moment, which is a source of GW, very like the binary. This GW is to be called 'continuous' GW (CGW), as it is emitted from the isolated object as long as it is magnetized and rotating along a misaligned axes. WDs, being very lowly luminous, pose a challenge to detect them via optical surveys like GAIA and SDSS. The super-Chandrasekhar WDs are supposed to be smaller, causing even lower luminosity. Possibly, for that reason, we did not detect any super-Chandrasekhar WDs directly. The detection of super-Chandrasekhar WD will become wide open via CGW once LISA is active. However, the GW amplitude will decay with time as the system radiates electromagnetic and GW energy by angular momentum extraction and slows down with time. The magnetic field also decays in different timescales due to the Hall and Ohmic dissipations. Together, those decay mechanisms provide a decay timescale in which the young WDs must be caught up by any detectors before the GW amplitude decays below sensitivity. Here, we aim to discuss the plausibility of detecting super-Chandrasekhar WDs via CGW. The plan of the paper is as follows. In section \ref{White dwarf model}, we discuss the deformed and misaligned rotating magnetized WD model, which is a viable candidate for GW radiation. We discuss the decay mechanisms responsible for GW amplitude decay in section \ref{GW decay}. We show the overall detection plausibility in section \ref{Detection plausibility} depending on the decay timescale. We finally conclude in section \ref{Conclusion}.

\section{White dwarf model}
\label{White dwarf model}

It is well known that any system is to be associated with non-zero time-varying quadrupole moment to generate CGW. The non-zero quadrupole moment can be generated from any fluid body with triaxial symmetry arising from deformation due to the magnetic field. The poloidal and toroidal magnetic fields provide a directional pressure, deforming the star into respectively an oblate and prolate shape. A rotating body with a misaligned axis and deformation will produce a non-zero time-varying quadrupole moment; the 'time-varying' part is because of rotation. A cartoon diagram of such rotating magnetized WD with misalignment between respective axes is shown in figure \ref{fig:cartoon}.
If the rotation rate of the WD is $\Omega$, it is well known that the body will radiate GW strain with two frequencies $\Omega$ and $2\Omega$ in two polarizations as follows \citep{BG1996, KM2019}:

\begin{figure}
\begin{center}
\includegraphics[scale=0.8]{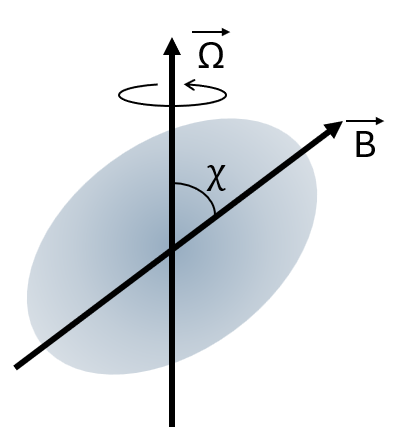}
\caption{A cartoon diagram of magnetized rotating WD
with misalignment between the magnetic field axis and the rotation axis.}
\label{fig:cartoon}
\end{center}
\end{figure}

\begin{equation}
\begin{aligned}
h_+ &= h_0 \sin \chi \left[ \frac{1}{2} \cos i \sin i \cos \chi \cos \Omega t - \frac{1+\cos^2 i}{2} \sin \chi \cos 2 \Omega t \right], \\
h_\times &= h_0 \sin \chi \left[ \frac{1}{2} \sin i \cos \chi \sin \Omega t - \cos i \sin \chi \sin 2 \Omega t \right].
\end{aligned}
\label{eq:gwstrain}
\end{equation}
Here, $h_0=\frac{2G}{c^4} \frac{\Omega^2 \epsilon I_{xx}}{d}\left(2\cos^2\chi-\sin^2\chi\right)$. 
For  $\chi \to 0$ (but $\neq 0$),

\begin{equation}
h_0=\frac{4G}{c^4} \frac{\Omega^2 \epsilon I_{xx}}{d},
\label{eq:gwamp}
\end{equation}
where $c$ is the speed of light, $G$ is Newton's gravitational constant, $\Omega$ is the angular frequency of the object, $d$ is the distance between the detector and the source object, $i$ is the angle between the rotation axis of the object and our line of sight, and the ellipticity, which is the degree of deformation, is defined as $\epsilon=|I_{zz}-I_{xx}|/I_{xx}$, where $I_{xx}$ and $I_{zz}$ are the principal moments of inertia of the star about $x$- and $z$-axes (any two orthogonal axes), respectively.
Further, the amplitudes of $h_+$ and $h_\times$ in equation (\ref{eq:gwstrain}) will be suppressed by an optimistic factor of $\left( \sin \chi \left[ \frac{1}{2} \cos i \sin i \cos \chi \cos \Omega t \right. \right. 
	\left.\left .- \frac{1+\cos^2 i}{2} \sin \chi \cos 2 \Omega t \right] \right)$, which can be maximized with respect to $i$ at $\chi=3^\circ$ (small obliquity angle or angle of misalignment) just during the birth of the WD, i.e., $t=0$, providing $h = 0.0110297h_0$, which we consider for further calculations.

\begin{figure}
\begin{center}
\includegraphics[width=\columnwidth]{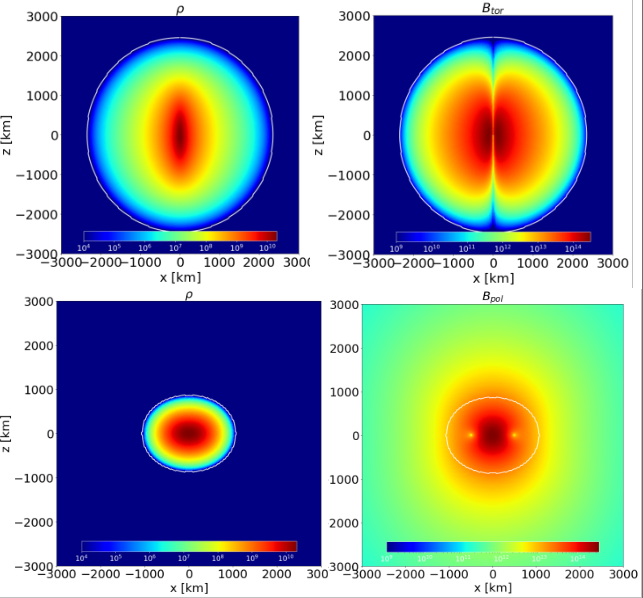}
\caption{Density isocontours (left) and magnetic field isocontour (right) for toroidally (top) and poloidally (bottom) dominated magnetized WDs of mass $M=2M_\odot$ and $1.6M_\odot$ respectively for toroidal and poloidal topologies with $B_{max}=3\times 10^{14}$ G. Magnetic to gravitational energy ratio $\sim0.1$}.
\label{fig:xns}
\end{center}
\end{figure}

In order to understand the plausibility of the detection of WDs via CGW, we have to calculate $h$, hence $h_0$, which depends on the precise measurement of quadrupolar deformation, hence the moment of inertia. To extract such information, we either need high optical resolution and observational data, which has not yet been achieved, or we need to model the WD with numerical simulation. We use a general relativistic steady state magneto-hydrostatic solver numerical code XNS 4.0 \citep{sold2021main} to model magnetized rotating stars. XNS 4.0 was originally designed for neutron stars, which we also modified to model WDs. We will design the WD interior with the relativistic equation of state for highly degenerate electron gas in the polytropic form $P=K\rho^{4/3}$, where $K=3.1\times10^{12}$ in CGS. We are showing some 2D crosssections and magnetic field topologies (toroidal and poloidal) here for the realization of the deformation (figure \ref{fig:xns}). It can be visualized that the WD with toroidal field is slightly prolate in shape, and the WD with poloidal field is oblate in shape. However, the angle between rotation and magnetic axes is zero here, and XNS cannot handle misalignment. Therefore, when we extract the ellipticity from XNS and use it as an input to calculate $h_0$, it is essential to consider a small misalignment so that the angle does not affect the physics. While extracting the ellipticity, we have to be cautious about only using the effect due to magnetic field deformation and rotation alone cannot provide nonzero quadrupole moment.

\section{Gravitational wave decay}
\label{GW decay}
\begin{figure}
\begin{center}
\includegraphics[width=\columnwidth]{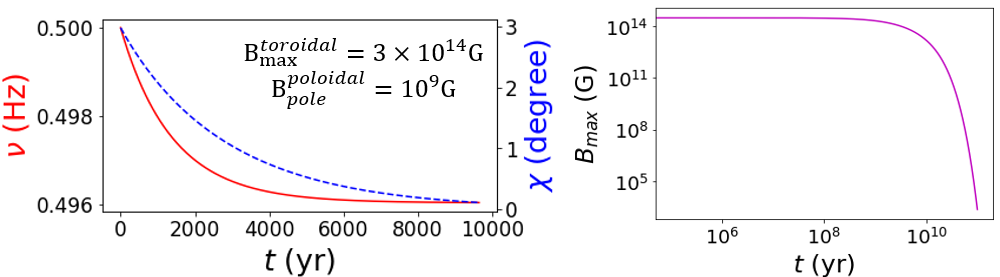}
\caption{Left: For toroidally dominated WDs, variations of $\nu$, $\chi$ as 
	functions of time for initial $\nu=0.5$ Hz, $\chi=3^\circ$. Red (solid) and blue (dashed) lines show the variations of $\nu$ and $\chi$, respectively. Right: Magnetic field decay of the same WD model for the maximum magnetic field $3\times 10^{14}$G.}
\label{fig:torwxlgw}
\end{center}
\end{figure}

 In section \ref{White dwarf model}, we have noticed that the GW amplitude is not stationary; rather, it is a function of $\Omega$, $\chi$, $\epsilon$, which in turn are functions of time during the evolution of WD. The ellipticity depends on the magnitude of the central magnetic field. A WD with a non-zero time-varying quadrupole moment will exhibit electromagnetic emission due to dipole radiation and gravitational radiation due to quadrupolar energy loss. Both of these energy losses lead to decay of $\Omega$ and $\chi$, meaning it slows down the star and tends to align the rotation and magnetic axes each other. The evolution equations are \citep{CH1970, das-mukhopadhyay} given by

\begin{equation}
\begin{split}
\frac{d(\Omega I_{z^{\prime}z^{\prime}})}{dt}=-\frac{2G}{5c^5} (I_{zz}-I_{xx})^2 \Omega^5 \sin^2 \chi (1+15 \sin^2 \chi)
-\frac{B_p^2 R_p^6 \Omega^3}{2c^3} \sin^2 \chi F(x_0)
\end{split}
\label{eq:dwdt}
\end{equation}
and
\begin{equation}
\begin{split}
I_{z^{\prime}z^{\prime}}\frac{d\chi}{dt}=-\frac{12G}{5c^5} (I_{zz}-I_{xx})^2 \Omega^4 \sin^3 \chi \cos \chi
-\frac{B_p^2 R_p^6 \Omega^2}{2c^3} \sin \chi \cos \chi F(x_0),
\end{split}
\label{eq:dxdt}
\end{equation}
where $x_0=R_0\Omega/c$, $B_p$ is the strength of the magnetic field at the pole, $R_p$ is the polar radius, $R_0$ is the average radius of WD, and the function $F(x_0)$ is defined as $F(x_0)=\frac{x_0^4}{5(x_0^6-3x_0^4+36)}+\frac{1}{3(x_0^2+1)}$. The set of equations (\ref{eq:dwdt}) and (\ref{eq:dxdt}) is to be solved simultaneously to obtain $\Omega$ and $\chi$ as functions of time. We need to supply $I_{xx}$, $I_{zz}$, $B_P$, and $R_P$ at $t=0$, which are the output of particular WD model from the {XNS} code. In the figure \ref{fig:torwxlgw}, the evolutions of $\Omega$ and $\chi$ can be visualized for toroidal field-dominated configuration, where the GW radiation, i.e., the quadrupolar term caused by deformation, plays a role in decay, and $B_P$ is negligible. The corresponding timescale of considerable decay before saturation occurs is around $10^4$ years. For poloidally dominated field configurations, the electromagnetic radiation plays an important role, and the decay happens much faster, in a couple of years! However, a stable star, here WD, should sustain a toroidally dominated field \citep{BR2009}. Hence, the former case is more realistic.
The spin and obliquity angle decay has been studied considering the magnetic field to be constant, which enforces $B_P$ and $I_{xx}-I_{zz}$ to be constant. However, this is not unrealistic, as the magnetic field decays in a much more longer timescale \cite{das-mukhopadhyay}. Nevertheless, eventually the field decays via the Ohmic decay and Hall drift. The Hall effect generally conserves magnetic energy. However, the larger turbulent eddies may be cascaded into smaller eddies when Ohmic decay will be dominated \citep{GR1992}. The timescales for the Ohmic and Hall effects are given as \citep{bhattacharya22} 

\begin{equation}
 t_{\text{Ohm}} = \left( 7 \times 10^{10} \, \text{yr} \right) \rho_{c,6}^{1/3} R_4^{1/2} \left( \frac{\rho_{\text{avg}}}{\rho_c} \right)   
\end{equation}
and
\begin{equation}
 t_{\text{Hall}} = \left( 5 \times 10^{10} \, \text{yr} \right) l_8^2 B_{0,14}^{-1} T_{c,7}^2 \rho_{c,10}   
\end{equation}
respectively,
where \( \rho_{c,n} \) is the central density in units of \( 10^n \, \text{g/cm}^3 \), \( R_4 \) is the radius of the star in units of \( 10^4 \, \text{cm} \), \( \rho_{\text{avg}} \) is the average density, \( \rho_c \) is the central density, \( l_8 \) is a characteristic length scale in units of \( 10^8 \, \text{cm} \), \( B_{0,14} \) is the magnetic field strength in units of \( 10^{14} \, \text{G} \), \( T_{c,7} \) is the core temperature in units of \( 10^7 \, \text{K} \). The magnetic field ($B$) decay in WDs can be studied by solving the decay equations \citep{HK1998}

\begin{equation}
\frac{dB}{dt}=-B \left( \frac{1}{t_{\rm Ohm}}+\frac{1}{t_{\rm Hall}} \right).
\label{eq:Bdecay}
\end{equation}
The initial magnetic field profile is taken from the {XNS} output data. We are showing the evolution of a magnetic field in figure \ref{fig:torwxlgw}, and the timescale of decay is more than 10 billion years, similar to the universe's age. This confirms that the calculation of spin down, where the magnetic field is considered constant, is correct in the sense that the magnetic field will remain constant indeed during spin down. In further studies of the complete decay mechanisms of GW, we will only take the effect of spin down into account as it is of the smallest timescale and, thus, more relevant. 

\section{Detection plausibility}
\label{Detection plausibility}
\begin{figure}
\begin{center}
\includegraphics[scale=0.8]{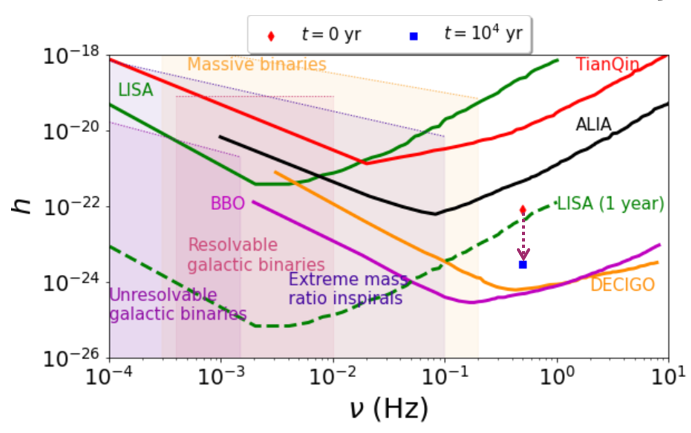}
\caption{Dimensionless gravitational wave amplitude before ($t=0$) and after ($t=10^4$ yrs) gravitational wave decay for a WD with $B_{max}^{Toroidal}=3\times10^{14}$ G, $B_{pole}^{Poloidal}=10^{9}$ G, initial $\nu=0.5$ Hz, initial $\chi=3^\circ$ and $d=5$ kpc.}
\label{fig:Bpoldecaygw}
\end{center}
\end{figure}

In this section, we will show the result of the decay mechanisms discussed in section \ref{GW decay} upon detecting GW by detectors. In figure \ref{fig:Bpoldecaygw}, a toroidally dominated WD with a magnetic field just after its birth and after the decay of $\nu/\chi$ due to GW emission over time is shown, assuming the distance of the object from the detector to be $5$ kpc. It is seen that the WD becomes undetectable by detectors after some time where it was detectable in the beginning. This sets an upper detection time limit, around $10^4$ years, which is still well out of the observational timescale (which can be a maximum of 5 years) and is good news. The question remains, however, in this `active timescale' when the WDs can be detectable in their first few thousands of years of lifetime, what is the probability of detection? To answer this question, we have to rely upon the birthrate of WD, which is $2\times 10^{-12}$ pc\textsuperscript{-3} yr\textsuperscript{-1} \citep{hol2016}. The birthrate in $5$ kpc volume is $0.25/yr$. Thus, around $2500$ WDs will be born in $10^4$ years in the nearest $5$ kpc volume from the Earth. All these WDs should be detected if they cross the minimum sensitivity of the upcoming detector LISA. However, the WD should be essentially highly magnetized to radiate CGW in the first place, and we know that $\sim 10\%$ of the isolated WDs are highly magnetized (e.g. \citep{Ferrario2020}), making the possible number of WD candidates to decrease to $250$. Now, it is well known that magnetized WDs are, on average, more massive than their nonmagnetic counterpart \citep{Ferrario2015}. However, no observation guarantees that all the highly magnetized WDs need to be more massive than the nonmagnetized ones. Hence, to target super-Changrasekhar WDs via CGW, we have to target the ratio of several overluminous Type Ia supernovae observed per year to standard Type Ia supernovae per year, which is about $2$\%. Moreover, there is no guarantee that the birthrate of WDs will be strictly proportional to the supernovae rate. However, one may expect the ratio of the standard WD to super-Chandrasekhar WD to be related to the correct percentage to consider when calculating how many WDs should turn out to be super-Chandrasekhar. Therefore, the number of super-Chandrasekhar WDs we can catch via CGW is around $5$ as soon as LISA starts operating.

\section{Method to enhance the detection possibility}
\label{Method to enhance the detection possibility}
\begin{figure}
\begin{center}
\includegraphics[scale=0.5]{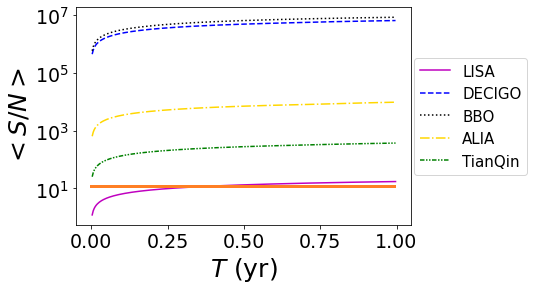}
\caption{SNR for various detectors as a function of integration time for the toroidally dominated WD (same as figure \ref{fig:Bpoldecaygw}) with initial $\nu=0.5$ Hz initial $\chi=3^\circ$ and $d=5$ kpc. The orange line corresponds to $<S/N>=12$.}
\label{fig:snr}
\end{center}
\end{figure}

The challenges of detecting WDs with LISA are to be appreciated. However, the probability of detection can be increased effectively by calculating the signal-to-noise ratio (SNR) cumulatively over a long observational time. The SNR in GW detectors measures how accurately a gravitational wave signal can be distinguished from the background noise in the detector. The calculation of SNR (S/N) is essential for determining the detectability of a GW event. If the GW strain remains constant over time, the SNR can be calculated as the ratio of the GW signal to GW noise as
\begin{equation}
    S/N=\frac{h_{signal}}{h_{noise}},
\end{equation}
where $h_{signal}$ is basically the GW strain of the object and $h_{noise}$ is a function of the detector's power spectral density ($S_n$). The SNR is to be considered constant for a GW instantaneously (for some fleeting seconds) emitted from merger events. However, we are dealing with CGWs, which require a `continuous' search over $T$ observational time. The SNR will change with time as the rotation rate and obliquity angle change. In this case, we have to discretize the total observational time into fine $\mathcal{N}$ time bins where the change of rotation rate and obliquity angle stays minimal. Then, we can carry out the time integration by simply summing over all the bins cumulatively. However, this addition is incoherent as the phase of GW for each time stack is different because the $\Omega$ changes with time (see \citep{MAG2008}, for details of the technique). Assuming $\nu$, $\chi$ and $h_0$ remain nearly constant over each time-stack, adding $\mathcal{N}$ such stacks, the cumulative SNR is \citep{das-mukhopadhyay}
\begin{equation}
\langle S/N\rangle =\sqrt{{\langle S/N_\Omega^2 \rangle}+{\langle S/N_{2 \Omega}^2 \rangle}}= \sqrt{\frac{\sin^2 \zeta}{100} \frac{h_0^2 T \sin^2 2\chi}{\sqrt{\mathcal{N}} S_n(\Omega)}+\frac{4 \sin^2 \zeta}{25} \frac{h_0^2 T \sin^4 \chi}{\sqrt{\mathcal{N}} S_n(2\Omega)}},
\label{eq:snr}
\end{equation}
where $\zeta$ is the angle between the interferometer arms, the data for power spectral density (PSD) of various detectors are extracted from https://gwplotter.com/. For ground-based interferometers $\zeta=90^\circ$ and space-based interferometers such as LISA $\zeta=60^\circ$. The SNR increases if we increase the observational time $T$; however, it decreases if we discretize the total observational time into more bins. Depending on how fast the spin down happens, we can minimize the number of bins so that the rotation rate and obliquity angle remain almost constant; here, we use $\mathcal{N}\sim300$. If we would have use $\mathcal{N}\sim 30$, the SNR would be $10^{1/4}\sim 2$. Taking care of all the hurdles during the long observational time for continuous search, we have considered the SNR threshold to be $12$, greater than the instantaneous search's threshold \citep{watt2008}. In the figure \ref{fig:snr}, we have shown a case where the WD would not be detectable immediately by LISA, but it can be detectable after couple months of integrated time. The SNR increases with time initially while the GW amplitude is added up in the first few time domains; however, with time, the decreased $\Omega$, $\chi$, and hence the decayed $h$ saturates the SNR with time. The slower the decay is, the more saturation is achieved later, opening the possibility of crossing the SNR threshold even in the later stage of observation. The slowly rotating WDs decay slower, providing the above-mentioned opportunity, though with lower GW emission in the first place. The WDs with higher magnetic fields are most likely to be detected as their $h_0$, hence SNR, is always higher.
The equivalent sensitivity curve, along with the sensitivity of many detectors for a $T$ time-long continuous search, is shown in the figure \ref{fig:snr} considering the sensitivity of the detectors is increased by $\sqrt{T}$ times as the continuous search is considered \citep{sold2021}, i.e.
\begin{equation}
    h\sim 11.4 \sqrt{\frac{S_n}{T}}.
\end{equation}
Using this technique, we have shown the instantaneous LISA sensitivity as well as the LISA sensitivity with 1 year of detection in figure \ref{fig:Bpoldecaygw}. In fact, the physics meaning of both the figures \ref{fig:Bpoldecaygw} and \ref{fig:snr} is the same.

\section{Conclusion}
\label{Conclusion}
We have aimed to discuss the detection plausibility of super-Chandrasekhar WDs via CGW with the upcoming GW detector in WD frequency LISA. By modeling the WD in general relativistic magneto-hydrostatics, we can understand what GW strain range is expected from the WDs, which are $5$ Kpc away from us. We also have pointed out that the GW amplitude is not constant with time; rather, it decays due to angular momentum extraction and magnetic energy loss. The effect of all the decays can be visualized in figure \ref{fig:Bpoldecaygw}. We have shown the result for one single WD model, which is toroidally dominated ($B_{max}^{Toroidal}=3\times10^{14}$G) with observationally relevant surface poloidal field ($B_{pole}^{Polooidal}=10^{9}$G). The poloidal field, however, is not strong enough to affect the structure of the WD. Hence, the GW amplitude is mostly from the quadrupolar deformation generated by the toroidal field. It is the most realistic model, as any stable WD should be associated with a toroidally dominated magnetic field with negligible poloidal field. In our model, the central toroidal field is at least $10^4$ times larger than that of central poloidal field strength. We also have discussed that a few super-Chandrasekhar WDs will be detected with upcoming detectors, keeping in mind the active timescale limit of the WDs. Although the hope for instantaneous detection can be almost ruled out by LISA, the exciting news is that LISA would be able to detect those super-Chandrasekhar WDs with 1 year of continuous detection, leading to direct detection of super-Chandrasekhar WDs for the first time, confirming their existence. 

\section{Acknowledgement}
MD takes this opportunity to thank BM for his support throughout her research so far. MD acknowledges the Prime Minister’s Research Fellows
(PMRF) scheme, with Ref. No TF/PMRF-22-5442.03.


\bibliographystyle{IEEEtran}

\bibliography{draft_m3}

 \end{document}